\documentclass{sn-jnl}% 

\usepackage[numbers]{natbib}

\usepackage[skipempty]{credits}

\usepackage{xcolor}
\usepackage{booktabs}
\usepackage{soul}
\usepackage{tikz}
\usepackage{adjustbox}

\definecolor{ACMBlue}{RGB}{1,130,172}

\newcommand{\myparagraph}[1]{\vspace{0.5\baselineskip}\noindent{\textbf{#1}}.~}
\newcommand{\myparagraphdot}[1]{\vspace{0.5\baselineskip}\noindent{\textbf{#1}}~}
\usepackage{lineno}

\usepackage{amsmath,amssymb,amsfonts}%
\usepackage{fancyhdr}
\hypersetup{
    colorlinks=true,
    linkcolor=ACMBlue,
    citecolor=ACMBlue,
    urlcolor=ACMBlue
}

\begin{document}

\thispagestyle{fancy}
\title{Control Search Rankings, Control the World: What is a Good Search Engine?}


\credit{Coghlan}  {1,0,1,0,1,1,0,0,0,0,0,0,1,1}
\credit{Chia}  {1,0,1,0,1,1,0,0,0,0,0,0,1,1}
\credit{Scholer}  {1,0,1,0,1,1,0,0,0,0,0,0,1,1}
\credit{Spina}  {1,0,1,0,1,1,0,0,0,0,0,1,1,1}

\author[1]{\fnm{Simon} \sur{Coghlan}}%[orcid=0000-0002-6021-9878]
\email{simon.coghlan@unimelb.edu.au}

\author[1]{\fnm{Hui Xian} \sur{Chia}}%[orcid=0000-0003-3075-3364]
\email{huixianchia@gmail.com}

\author[2]{\fnm{Falk} \sur{Scholer}}%[orcid=0000-0001-9094-0810]
\email{falk.scholer@rmit.edu.au}

\author[2]{\fnm{Damiano} \sur{Spina}}%[orcid=0000-0001-9913-433X]
\email{damiano.spina@rmit.edu.au}

\affil[1]{\orgname{University of Melbourne}, \orgaddress{\city{Melbourne}, \state{VIC}, \country{Australia}}}
\affil[2]{\orgname{RMIT University}, \orgaddress{\city{Melbourne}, \state{VIC}, \country{Australia}}}

\abstract{

This paper examines the ethical question, `What is a good search engine?' Since search engines are gatekeepers of global online information, it is vital they do their job ethically well. While the Internet is now several decades old, the topic remains under-explored from interdisciplinary perspectives. This paper presents a novel role-based approach involving four ethical models of types of search engine behavior: Customer Servant, Librarian, Journalist, and Teacher. It explores these ethical models with reference to the research field of information retrieval, and by means of a case study involving the COVID-19 global pandemic. It also reflects on the four ethical models in terms of the history of search engine development, from earlier crude efforts in the 1990s, to the very recent prospect of Large Language Model-based conversational information seeking systems taking on the roles of established web search engines like Google. Finally, the paper outlines considerations that inform present and future regulation and accountability for search engines as they continue to evolve. The paper should interest information retrieval researchers and others interested in the ethics of search engines.
}

% Keywords

\maketitle

\section{Introduction}
Search engine (SE) results substantially affect what information the public can see and access on the internet and how that information presents to them. This control of the vast information on the internet gives popular SEs like Alphabet's Google and Microsoft's Bing great ability to influence beliefs and opinions, and cause individual and societal benefit and harm. It is therefore important that SEs do their job as ``gatekeepers'' \citep{hinman_esse_2005} of the world's online information well. But what exactly, in a broad ethical sense, is a ``good'' web SE? Although some ethics-related papers have considered the qualities of a good recommender system or SE \citep{introna2000shaping,taddeo2016debate,tavani2020search,hinman2008searching,granka2010politics,goldman2008searcha,polonioli_ethics_2021,tang_i_2016, hinman_esse_2005, munton2022answering,heersmink_virtue_2018,elgesem2008search}, and there have lately been several critiques of, e.g., Google \citep{mager_advancing_2023,graham_ethical_2023,noble2018algorithms}, there are still relatively few truly interdisciplinary analyses \citep{gao_toward_2020,shah2022situating,shah2024envisioning}. Indeed, despite the World Wide Web (WWW) now being several decades old, the important question of how SEs ought to -- and also how they can -- operate remains under-explored from interdisciplinary perspectives \citep{spink_web_2008}. In this paper, we combine ethical discussion with information retrieval (IR) knowledge, plus some regulatory reflections, to address the question of what a good search engine might look like. The analysis will help cross the disciplinary divide between ethics and IR and inform interested parties from both disciplines.

\myparagraph{Scope}

Before we begin the examination, some clarification of its scope is required. We shall focus on so-called ``horizontal'' SEs that retrieve and present information from the WWW, such as Google, Bing, and DuckDuckGo, rather than on more ``vertical'' systems that return information and recommend listings concerning (say) movies, job positions, music, videos, or other narrowly defined materials -- although vertical systems also raise ethical issues \citep{alfano_technologically_2021, seaver_captivating_2019}. We also concentrate on internet search related to varieties of information (e.g., politics, news, history, science, art, etc.) rather than on online activities and services such as retail, banking, and social media. While we take an ethical perspective, we bypass a range of internet and SE-related ethical issues \citep{tavani2020search,spink_web_2008,frost-arnold_trustworthiness_2014} such as user and data privacy and the right to be forgotten \citep{taddeo2016debate}.

In questioning what a good SE is, we are not asking how SE companies might best serve their customer businesses (e.g., via paid ads), enhance surveillance of users \citep{zuboff_surveillance_2019}, and maximize company profit. Rather, we are asking what a good SE might be from the perspective of other parties, particularly information seekers and human society generally -- though there could also be ethically important implications of SEs for non-human animals and the environment too \citep{coghlan_harm_2023,hagendorff_speciesist_2023}. 

We are aware that a comprehensive approach to answer ethical questions about how to design SEs may require SE companies to pay attention to factors like profit and shareholder value. Nevertheless, we shall quarantine these considerations from our analysis, as we are interested in characterizing what a good SE might look like in a broader sense that is independent of a given SE company's own interests. I.e., our work concentrates on the ethical implications of technological/algorithmic aspects of SEs rather than in business operations. Even so, we believe a better understanding of what a good SE is may inform SE companies -- as well as regulators and the public society -- to make more ethically informed decisions about appropriate SE design and use.
\newpage
\myparagraph{Paper Contribution}

This paper addresses the gap in interdisciplinary research between the very different fields of ethics and IR regarding the operation and role of SEs. The field of IR has developed a wide array of methods to find information that a user is seeking, and to evaluate SE performance, but there has been relatively limited ethical consideration there about how SEs may affect users and society \citep{ekstrand2022fairness,ekstrand2024not,hofmann2016online,milano_recommender_2020}.
And while there has been some philosophical work on virtue and epistemology issues with SEs \citep{simpson2012evaluating,munton2022answering,narayanan2022google,heersmink_virtue_2018}, there is a relative gap concerning how ethical analyses translate to concepts in Information Retrieval (IR) and whether and how ideals are already and may later be operationalized in actual systems. Likewise, a gap exists in the field of IR concerning how decisions regarding search optimization reflect value judgments and assumptions.

To this end, we propose a novel role-based approach that draws on medical ethics to articulate four models conceptualizing the role of SEs: (i) \emph{Customer Servant}, (ii) \emph{Librarian}, (iii) \emph{Journalist}, and (iv) \emph{Teacher}. Notwithstanding some qualifications that we will outline, these four models provide a useful foundation for ethical discussions about how a SE should behave in various scenarios -- for example, when harm to society might arguably justify manipulation or censorship of information by a SE. The models of SEs are then linked to different components and technologies developed in IR, providing the groundwork to bridge the divide between ethical questions of how SEs might behave and techniques in IR.

How the four ethical models might be used by interested parties and stakeholders to reason and argue about what makes a ``good'' SE is illustrated in a discussion of web search during the COVID-19 pandemic. Our approach there is not to argue definitively for any particular SE model, but instead to inform ethical, legal, and IR debates by clarifying important ways SEs might operate in the provision of information. This provides a foundation for the development of policies and regulations of information access and retrieval online—a vital ongoing issue for society and law. A key challenge for effective regulation of SEs is the gap between principle and practice -- laws impose value judgments at the level of abstract principles, but how that translates to actual implementation in SEs remains uncertain. Our approach bridges the gap between principles and implementation, explaining how different SE designs are consistent with different value judgments.

As we noted, SE technology has evolved over the decades and is likely to further evolve. Thus, toward the paper's end we discuss the possible next generation of SEs, namely conversational search driven by large language models (LLMs) such as GPT-4. Since LLMs may radically alter the nature of online search, it is timely and helpful to discuss them in terms of our ethical SE models. 

The rest of the paper is organized as follows. Section~\ref{sec:search} outlines how search engines work and are evaluated in the field of IR. Section~\ref{sec:models} presents four ethical models for SEs and examines how they are implemented in existing systems. Section~\ref{sec:covid} offers an exploratory COVID-19 case study to illustrate application of the four models. Section~\ref{sec:discussion} discusses accountability, the emergence of powerful chatbot-style SEs, as well as authors' positionality and ethical considerations of our work. Section~\ref{sec:conclusions} concludes.

\section{Search Engines}
\label{sec:search}

\subsection{How Search Engines Work}
\label{sec-se}

A SE is a type of IR system, a software tool that aims to help a user to satisfy an \emph{information need} \citep{belkin1982ask}. SEs were developed specifically to enable finding information on the WWW. From a \emph{system} perspective, a SE conceptually consists of four core components \citep{croft_search_2010}: (i) a \emph{crawler} which finds documents on the Web and adds them to a local database or \emph{collection}; (ii) an \emph{index} which supports the fast finding of individual documents in the collection based on their content (e.g., the individual terms that occur in a text document); (iii) a \emph{ranker}, which creates an ordering of the documents in the collection based on their likely \emph{relevance} to a user’s \emph{query}, following the Probability Ranking Principle \citep{robertson1977probability}; and (iv) an \emph{interface} though which the user can submit their search request to the system. 

From a \emph{user perspective}, a SE is used to resolve an information need. The key task for the user is to translate their information need into a specific form that is interpretable by the SE, typically by writing a \emph{query} that consists of individual \emph{terms} (or keywords). The SE responds by presenting a ranked list of documents as search results, with those that the system estimates to be more likely to be relevant being ordered above those that are estimated to be less likely to be relevant. IR is challenging for a range of reasons. Language is complex due to features such as \emph{synonymy} (multiple words being used to refer to the same concept), and \emph{polysemy} (the same work having multiple meanings. 

Moreover, user information needs can be highly diverse \citep{broder2002taxonomy}, ranging from re-finding a resource that the user has seen before (e.g., homepage of the New York Times), to helping the user to learn about a topic that they know little or nothing about (e.g., agrarian society during the Middle Ages). Note that information needs can vary in many dimensions, including complexity (finding a single resource, versus learning about a complex topic), as well as specificity (finding something that the user is able to clearly describe or already knows, versus finding something that is largely unknown to the user and so may even be difficult to conceptualize and explain \citep{belkinASKInformationRetrieval1982}). Successfully finding useful information is therefore a human as well as a computational challenge.  

The results the system presents to the users are most directly determined by the SE ranker component. Traditionally, SEs estimate the relevance of documents based on term matching, or the co-occurrence of terms that occur in a query submitted by a user with terms that occur in a document. This is further refined by statistical considerations of term distributions in the collection of documents \citep{jones_probabilistic_2000}: \emph{term frequency}, whereby documents that include a greater number of query terms are more likely to be ``about'' the same topic as the query; and \emph{inverse document frequency}, whereby those terms that occur in fewer documents across the collection are more discriminative, and should therefore be treated as more important for estimating relevance. Rankers can also include other sources of evidence, including features such as the \emph{authoritativeness} of documents (e.g., informed by the document’s authorship, or standing of an institution); \emph{popularity} (e.g., Google’s PageRank upweights documents that include more incoming links from other documents by considering the links as an indication of the document’s importance); \emph{timeliness} (e.g., the age of a document); and other features learned using statistical or machine learning processes. These derived features are then combined with the term weighting information to obtain an overall ranking of documents in response to the submitted user query.

\subsection{Evaluation of Search Engines in Information Retrieval}

In IR, the success of a SE can be conceptualized in different ways, ranging from \emph{user satisfaction} (how happy is the user with the system’s ability to help them to resolve their information need?), system \emph{efficiency} (how quickly were the results returned to the user?), or even \emph{revenue} (how much money was made by showing and enticing a user to click on ads that were included on the search results page?). The most widely used approach for measuring effectiveness is through offline evaluation with the use of test collections \citep{sanderson2010test,harman2019information}, where a pre-determined and representative set of test queries are run using a SE, and each of the returned answer documents is independently judged regarding its \emph{relevance}. Relevance is itself a complex, contested, and nuanced concept \citep{mizzaro1997relevance,mizzaro1998how,munton2022answering}, but is typically operationalized into a consideration of ``aboutness'': is this answer document about the same topic as the query? 

The relevance assessments of individual answer items in a list of ranked search results are aggregated into different measures, each of which instantiates a specific user model. For example, \emph{Precision@10} counts how many relevant documents occur in the first 10 positions of the ranked answer list, with a corresponding user model of a searcher who only looks at the first page of search results. The popular \emph{Normalized Discounted Cumulative Gain (NDCG)} measure \citep{jarvelin_cumulated_2002} includes the notions of \emph{degree} of relevance (some documents are more relevant than others, and should therefore contribute more to the system’s effectiveness score), and \emph{discounting} the utility of documents the lower their position in the ranked list (since it takes increasing levels of time and effort the longer that a user works their way down the results list, relevant items that are placed higher in the ordered results list should contribute more towards the system’s effectiveness). Choosing a test collection and an effectiveness measure then allows for the comparison of different SEs, in the sense that system A may score higher than system B (with the level of confidence usually further evaluated using a statistical significance test).

\subsection{Search Engines as Providers of Information}

While offline evaluation using test collections, together with online evaluation based on search logs and implicit user feedback \citep{hofmann2016online}, has enabled remarkable advances in the performance of SEs over the last decades, particularly from the perspective of the ranking algorithms and features that identify relevant results, it is clear that this approach does not incorporate many other features, including ethical ones, that could be taken into account when considering systems that determine which \emph{information} users across the globe see in response to the billions of queries that are submitted on a daily basis. In fact, web SEs have a large impact in influencing how we consume information online, and can influence the way searchers see the world \citep{hinman2008searching,nguyen2023ai}. This is evidenced by recent regulation initiatives, such as the EU Digital Services Act \citep{eu2022dsa}.

The essential task a SE carries out is to retrieve information for the user that helps them  resolve an information need. The essence of this task is deciding which information (documents) should be shown, and moreover, which ones should be more prominently displayed by being placed higher in the ranked results list. The ranking has a key impact on the delivery of information, since studies that have analyzed user behavior (using features such as eye tracking \citep{buscher09}, and implicit user signals such as mouse clicks on items in the results list \citep{chuklin2015click}) have repeatedly shown that most attention is devoted to the single top-ranked item on a search results page, with rapidly decaying levels of attention being given to subsequent items. On average, few users continue beyond the last item that is displayed when a search results page is first displayed (typically around 6 or 7 items for desktop computer displays, and fewer for mobile displays) \citep{moffatUsersModelsWhat2013a}. Some users will continue by scrolling down to the bottom of the first results page, but even fewer then move on to a subsequent page of results (by default, most SEs display 10 items per page). The practical implications of a document’s position in a SE’s answer page are profound.\footnote{As SEs became increasingly essential for finding information online, a new SE optimization industry sprang into being, businesses whose purpose is to help other companies or individuals to improve the rank position at which their information will be placed on a search results page.}  

SEs rely on a ranking algorithm to determine the order in which items are presented to users, as explained above. Conceptually, these ranking can be determined fully automatically in response to any submitted user query, with no manual human intervention in the produced rankings (i.e., without a human editor who chooses to alter the positions of certain items, or to remove them entirely from a results list). However, the ranking algorithms themselves are tuned and evaluated, based on labeled training data. For example, a key notion is the assessment of the \emph{relevance} of returned answer documents to the query for which they were returned, as explained above. These assessments are made by humans, following certain guidelines. For example, Google has released a 168-page document containing detailed instructions about the features that assessors should take into account when evaluating search results \citep{google2023quality}. SEs will therefore incorporate certain biases, by design, based on the criteria by which their results are judged (and how these are interpreted and put into practice by individual human assessors), which are then in turn used to tune the SE to return the ``best'' documents at higher rank positions.

\section{Four Ethical Search Engine Models}
\label{sec:models}

We now turn to the broader ethical question of what exactly a good SE might be. We present a role-based approach consisting of four models of how SEs might operate, inspired by work from medical ethics. In a highly influential paper, \citet{emanuel1992four} describe several models of the doctor-patient relationship. The authors' aim in articulating these models for doctors was to lay out clearly various possible normative relations between doctors and patients. Once the possibilities were distinguished and described, the taxonomy could serve as a basis for reflection and argument about how doctors \textit{ought} to interact with patients. This was meant to assist both medical personnel and ethicists. 

Now, the \textit{paternalistic} doctor, say \citeauthor{emanuel1992four}, seeks to promote the patient’s health even when it conflicts with the patient’s choices. For example, the paternalistic doctor may present just that information that will incline the patient to choose the intervention the doctor deems `right'. The \textit{informative} doctor, by contrast, simply presents truthful information to the patient and lets them decide what to do. 

The \textit{interpretive} doctor goes further and seeks to interpret the patient’s autonomous wishes regarding health interventions. Autonomy is often understood as a mature person’s ability to self-govern: to live their own life according to their own values free of paternalistic or other interference \citep{young2017personal}. The interpretive doctor presents patients with facts relevant to their autonomy. Because a patient’s values may be unclear, the doctor may try to elucidate them, without otherwise seeking to influence patient choice. 

The \textit{deliberative} doctor goes further still, discussing with the patient values associated with different intervention options and sometimes aiming through dialogue (not coercion) to sway or convince the patient of the rightness of certain interventions. This doctor is more like a ``teacher or friend'' \citep[p.~2222]{emanuel1992four}. Finally, the \textit{instrumental} doctor aims at goals beyond the patient, such as societal good, even at the expense of the patient's interests and wishes.

Importantly, these models of the role of doctors are not purely descriptive but also normative. Each highlights potentially ethically significant qualities in doctors that help us reflect on relevant yet contestable kinds of interaction. Our SE models have similarities and differences with the medical models. Like those medical models, they are not exhaustive and are simply intended as ‘ideal types’ 
\citep[p.~2221]{emanuel1992four} that illustrate contrasting but important possibilities. To avoid initial distractions about the feasibility of implementing these models in existing SEs (which are predominantly owned and run by for-profit corporations), we imagine constructing these SEs from scratch. These models assume search results can be presented in various ways. We draw explicit links between them and existing IR technologies in Section~\ref{sec-implementing} below.  

The four role-based models we now describe are (i) \emph{Customer Servant}, (ii) \emph{Librarian}, (iii) \emph{Journalist}, and (iv) \emph{Teacher}. Each model explains SE behavior in terms of an analogy with familiar human occupations. To avoid misunderstanding, we must emphasize that the analogies are not supposed to be perfect: inevitably, there will be differences between how humans in those occupations, and how the SEs that have resemblances with them, perform their respective roles. For example, a SE that behaves roughly analogously to a human librarian is not like a librarian in all respects. We also somewhat simplify the occupations that form the basis of the analogical SE models. Nonetheless, these SE models enable various parties and stakeholders (including those who lack a grasp of technical details) to more readily understand ethically salient features of SEs.

\paragraph{Model 1: Customer Servant.} In retail, a customer service agent is motivated to give customers what they ask for, even if that is not what the agent believes they really want or is good for them. If the customer asks for a certain type of white good or food product, for example, the customer service agent might simply direct them to those items, without necessarily asking any further questions or making judgments about what the customer might actually value or really prefer (even though, of course, some customer service agents will do that in real life). 

Put in terms of a SE, the \textit{Customer Servant }algorithm returns only those results which are germane to the specific query as it has been articulated by the user on that precise occasion. \textit{Customer Servant }somewhat resembles the informative doctor who is content to provide information based simply on a person’s most recent question or request, without probing any deeper for a better understanding of the patient’s intentions and their underlying values and goals.

\paragraph{Model 2: Librarian.} The human librarian tries to give information searchers the resources they are truly seeking. Unlike the retail customer service agent, the librarian may draw on additional evidence or information to infer the searcher’s actual intentions which may not be fully represented in their queries and requests. For example, the librarian may seek further clarification by asking additional questions, and give a range of options they think may reflect the borrower's interests. Because the librarian can `get to know' the borrower, they may offer suggestions based their habits and previous reading based on their loan history. For example, the librarian who knows that a reader likes engineering, or conspiracy theories about aliens and moon landings, may make corresponding recommendations. Librarians do not just `blindly' respond to specific queries.

In SE terms, this algorithmic agent takes note of the specific query but also uses other information such as context (location, time, or query history) to infer the user’s search goals and intentions. By using the search query plus other data, \textit{Librarian} seeks to provide information that is as relevant as possible to the user’s actual search goals \cite{munton2022answering}. The Librarian SE model somewhat resembles the informative and interpretative doctor models. Unlike the deliberative doctor, \textit{Librarian} is, in a certain sense, `value neutral'. That is, the analogy here is with the kind of human librarian who does not interfere in any way with the borrower's request. 

In reality, of course, librarians may make personal recommendations about books and other materials, and they may attempt in various ways to sway the information seeker. Human librarians may also exercise discretion over which information and texts enters the library in the first place. But for the purposes of characterizing our \textit{Librarian} SE model, we are imagining a human librarian who simply endeavors to work out what the person is seeking and to provide that information without question or judgement. 

\paragraph{Model 3: Journalist.} In contrast to librarians (as we have characterized them), the professional journalist not only attempts to provide the information people are actually seeking, but also information that helps them achieve a reasonable, if basic or rudimentary, understanding of the relevant topic. This requires presenting arguments and facts (not falsehoods or misleading information), typically in a somewhat balanced way. For example, even if many readers or viewers are interested in conspiracy theories, the journalist might include debunking information. Similarly, if someone requests information about war crimes from one side of a conflict, the journalist might also present information about war crimes perpetrated by the other side \citep{simpson2012evaluating}. The journalist may also avoid endorsing or appearing to endorse social prejudices like racism, sexism, ableism, and ageism, and avoid excluding perspectives from minoritized groups \cite{castillo2019fairness}

While the librarian (as we characterized them) eschews an ethically value-laden position (except perhaps for withholding illegal materials), the journalist, as a member of the `fourth estate', sees themselves as promoting social good to some extent, mostly by ensuring that citizens are adequately informed in a basic way while also promoting free speech and avoiding the perpetuation of social prejudices. To meet these goals, the journalist frequently draws on—while often greatly simplifying for their readership—reputable sources and expert knowledge.

In SE terms, this agent satisfies not only the IR goal of `relevance' but also of aiming to educate users to some basic degree about the facts and the truth while avoiding prejudice and unfair exclusion \cite{noble2018algorithms, whitney_simpson_2019}. Roughly speaking, \textit{Journalist} has elements of the deliberative and instrumental doctor models, since it responds not only to user intentions but also to its 'conception' of what might be good for society.

\paragraph{Model 4: Teacher.} A teacher is supposed to have deep knowledge of their subject and the ability to make fine judgments about that subject matter. Teachers aim to go beyond merely transmitting knowledge to acting in the best interests of their students as learners and individuals, even if they do not always succeed in that endeavor. Because students exposed to many opinions may not readily discern truth and acquire genuine knowledge, teachers generally try to guide students to truth and deeper understanding while encouraging critical thinking. If the student reads bad history or pseudoscience, the teacher may either debunk those views with their specialist expertise, or even deter such reading altogether.

Teachers thus go a step beyond journalists. For example, while journalists partly cater to readers' interests in being entertained or amused, teachers focus on more profound edification and often prefer intellectually difficult over ``dumbed-down'' material. Moreover, the teacher aims not only to impart knowledge and intellectual virtue (e.g., critical thinking skills and judgment), but often also ethical virtue \citep{vallor2016technology}. For instance, a teacher may actively \textit{discourage} a student from reading works of holocaust denial which are not only pseudohistorical but potentially morally corrupting. Furthermore, teachers aim to benefit society in ways that go well beyond the social service that journalists aim at, by producing students with rich and socially useful knowledge who can become morally decent or exemplary citizens.

In the \textit{Teacher} SE model, it is presumed justified to almost always give users what it 'judges' or 'deems' correct information. \textit{Teacher} also aims to give substantial answers to search queries while severely demoting or hiding results that, though entirely relevant, are judged incorrect, harmful, or 'merely' superficial. \textit{Teacher} is in a sense the most paternalistic model, as it assumes users may be incapable or less capable of deciding what information to seek and believe in their own or society’s best interests. Like human teachers, this SE model may not always (or even, depending on the case, often) succeed in educating and edifying as opposed to misleading and corrupting. Nonetheless, \textit{Teacher} can be regarded (in the spirit of the analogy) as \textit{attempting} to inform, guide, and enlighten \cite{metzler2021rethinking}.

\subsection{Remarks on the Four Models}

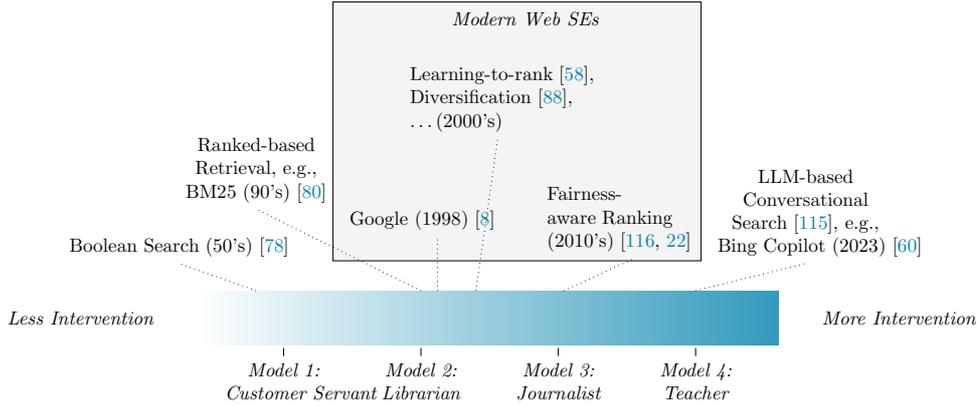
\begin{figure}[tp]
\centering
\adjustbox{max width=1\textwidth}{
\begin{tikzpicture}
\tikzset{
    mynode/.style={
      text width=3.2cm, % Set the maximum width for text
      align=center,   % Enable text alignment
      color=black
    },
    modelnode/.style={
      text width=2.14cm, % Set the maximum width for text
      align=center,   % Enable text alignment
      color=black
    },
    node/.style={
    color=black
    }
  }

\shade[left color=white,right color=ACMBlue!80] (0,0) rectangle (10.5,1);

\draw[fill=gray!9] (2.4,1.55) rectangle (9.1,6.3);  
\node[align=center] at (5.9,6) {\it \color{black}{Modern Web SEs}};

\draw (-1.2,0.5) (-2.2,0.5) node{\it \color{black}{Less Intervention}};
\draw (1,0.5) (12.7,0.5) node{\it \color{black}{More Intervention}};
\foreach \x/\y in {1/{\it Model 1: \mbox{Customer Servant}},2/{\it \mbox{Model 2:} \mbox{Librarian}},3/{\it \mbox{Model 3:} \mbox{Journalist}},4/{\it \mbox{Model 4:} \mbox{Teacher}}}
\draw (\x*2.5-1,-0.05) -- (\x*2.5-1,-0.2) node[modelnode,below] {\y};

%Model 1: 1.5
\draw[dotted] (1,1) -- (-0.4,1.5) node[above,align=left] {\color{black}{Boolean Search (50's)}~\citep{rijsbergen_information_1979}} ;

%Model 2: 4

\draw[dotted] (4,1) -- (1,2.5) node[mynode,above] {Ranked-based Retrieval, e.g., BM25 (90's)~\citep{robertson1995okapi}};
\draw[dotted] (4.3,1) -- (4.3,2) node[mynode,above,align=left] {Google (1998)~\citep{brin1998anatomy}};
\draw[dotted] (5,1) -- (5.4,3.8) node[mynode,above,align=left] {\mbox{Learning-to-rank~\citep{liu2007learning}}, \mbox{Diversification} \mbox{\citep{santos2015search}},\\\ldots(2000's)};

%Model 3: 6.5
\draw[dotted] (6.6,1) -- (7.9,1.6) node[mynode,above,align=left] {Fairness-aware Ranking (2010's)~\citep{10.1145/3533379,ekstrand2022fairness}};

%Model 4: 9
%\draw[dotted] (9,1) -- (9.5,3.2) node[mynode,above]  {Spam Filtering \citep{10.1145/1017074.1017077}};
\draw[dotted] (9,1) -- (11, 1.5) node[mynode,above]
{LLM-based \mbox{Conversational} Search~\mbox{\citep{zamani2023conversational}}, e.g., \mbox{Bing Copilot (2023)}~\mbox{\citep{mehdi2023reinventing}}};

%ChatGPT (2022) \citep{brown2020languange,openai2023gpt4}};
\end{tikzpicture}
}

\caption{\label{fig:models} A schematic representation of the relation between the four ethical models, the degree of intervention -- both inferential (related to guessing relevance) and ethical  (related to what is considered good for the user and/or society) -- that models take into account, and examples of instances of IR techniques throughout that spectrum.}
\end{figure}

Models and analogies can be both useful and misleading, including in relation to SEs \cite{whitney_simpson_2019}. To help avoid misunderstanding of this four model role-based approach, we make the following remarks. First, the four ethical SE models need not be mutually exclusive. As with \citeauthor{emanuel1992four}’s medical models \citep[p.~2225]{emanuel1992four}, there may be some occasions, but not others, on which a particular model is justified (we discuss this below) \cite{shah2022situating}. Second, each SE model has pros and cons, which may require trade-offs in practice. Third, SEs in reality may contain elements of more than one model. For example, they may sometimes operate more like \textit{Librarian} and other times more like \textit{Journalist} -- although one type may generally predominate. Fourth, the four models exist along a spectrum of two main countervailing values (Figure~\ref{fig:models}), from meeting user wishes, to a more paternalistic goal of promoting users' or others' interests. We discuss this matter further in our case study (Section~\ref{sec:covid}).

Fifth, while the models represent `ideal types' from different perspectives, there may be real-world circumstances that prevent the SE from doing a first-rate job as the type of model it is. For example, as just alluded to, \textit{Teacher} may provide less than the best and most expert information about a given topic, while \textit{Journalist} may provide less than fully-balanced information about a particular issue. This could be due to limitations in the material available or accessible on the internet \cite{barabasi2002linked,munton2022answering}. A similar problem could also occur when the search process involves a Large Language Model (LLM) that has biased, incomplete, or missing information in its training data. We discuss LLMs in more detail in Section~\ref{sec-LLMs}. Finally, since we are interested in models that have something going for them in an ethical sense (even if no model is ethically perfect and suited to all circumstances), we focus on modeling the organic search results of SEs and do not include other possible types, such as a model that merely seeks to advance a (short-sighted \cite{elgesem2008search}) SE company's advertising interests.\footnote{If we did include such a model, it might be called, say, \textit{Corporate Lackey}. We leave this for future work.}

\subsection{Implementing Ethical Models in Search Engines}
\label{sec-implementing}

% A handy docuement by Mark and Bruce: https://ciir-publications.cs.umass.edu/getpdf.php?id=1066

We now consider ways the ethical or normative models can be instantiated, drawing on IR technologies and research.

\emph{Model 1 (Customer Servant)} is based on the principle of directly returning what the user requested, with no further inference. This is most closely matched by Boolean IR systems, the primary retrieval paradigm before the notion of ranked retrieval became popular \citep{sanderson2012history}.  Boolean search is based solely on keyword matching, with individual query terms being modified through the use of of Boolean operators AND, OR, and NOT \citep{rijsbergen_information_1979}. A Boolean query returns a set of documents (i.e., an unordered group of documents) that formally match the criteria that the query specifies. For example, the query \texttt{pet AND food AND cat AND NOT fish} would return every document that includes the three terms pet, food, and cat, while specifically not including those documents that also include the term fish. A key property of Boolean retrieval is that any items that occur in the result set can be directly related back to the query, leading to a highly explainable approach for how information is obtained; this retrieval model is therefore commonly deployed in search systems that are used by domain experts, for example in medical literature search, library search, or the process of e-discovery that is carried out as part of legal proceedings. While popular modern SEs typically support the notions of Boolean retrieval as ``advanced search'' features, these operators are very rarely used \citep{spink_searching_2001}, and the retrieved documents are still ranked based on various other features, rather than being returned as an unordered set. Therefore, \textit{Customer Servant }can be practically instantiated and is important in certain domains, but is not widely used by web SEs.

\emph{Model 2 (Librarian)} most closely reflects the behavior of current web SEs on the widest spectrum of queries: in addition to matching the component terms of the user's current query, other evidence is used to try to better infer what the user is actually seeking. For example, user interactions and behavioral logs \citep{white2016interactions} (historical clicks of other users who have previously submitted the same query) are one of the most informative sources to estimate ``relevance'' for already seen queries (e.g., a link to ``www.facebook.com'' is returned as the top search result for the query ``facebook'', because in the past most users clicked on that result item when they submitted the query; based on term matching alone, the web page at www.facebook.com might not be the ranked as the best match). The goal of \emph{learning-to-rank} \citep{liu2007learning}  documents by combining a number of features beyond pure term matching, typically using machine learning, is to reduce the semantic gap between the direct query terms and the \emph{meaning} of those terms, both in relation to the information need and the content of documents as intended by their creators.

Further mechanisms that modern SEs deploy to assist in ranking answer documents by inferring more than is available simply from term matching include advising the user of related searches, and providing query auto-completion mechanisms to assist the user when specifying their information need \citep{10.1145/2600428.2609508}. Ambiguity is addressed via search results diversification and contextualization (e.g., by using location and time as context features to tailor results to the specific user). 
Clarifying questions can be used when the system has low confidence in the effectiveness of the search results, asking the user to further expand on their information need specification \citep{10.1145/3366423.3380126}.
Another enhancement that has contributed to effectiveness by better inferring the meaning behind a user's query is the use of dense representations of terms using embeddings from pre-trained language models, which enable matching on semantic concepts, as opposed to matching directly on terms themselves \citep{10.1145/2766462.2767780}. 

Importantly, the described approaches are deployed algorithmically -- while some of them may have been tuned based on human editorial judgments in a broad sense, there is no notion of the ``organic'' search result rankings, once generated by the system, being further edited through direct and targeted individual intervention. Modern SEs therefore currently best align with \textit{Librarian} for the majority of searches.

To align with \emph{Model 3 (Journalist)}, a practical SE would need to employ additional actions to alter the ranked list of answer documents that are returned algorithmically based primarily on the notion of their relevance (as for \textit{Librarian}). In fact, current SEs do instantiate this model for a small portion of the query spectrum -- for example, Bing introduced in 2017 multi-perspective answers for a number of sensitive topics such as health-related queries (e.g., ``is hot yoga good for you?'') \citep{bing2018multiperspective}. Head queries (the most frequent queries received by a web SE over a period of time) are more likely to receive these type of interventions, including viewpoint diversity, e.g., to mitigate confirmation bias \citep{draws2023viewpoint}; providing additional context (e.g., from the knowledge graph) to give more agency to users~\citep{google2019how}; balancing or diversifying content and suggestions (e.g., related searches) to mitigate the amplification of stereotypes and discrimination \citep{noble2018algorithms} (e.g., balancing image results in terms of gender and ethnicity for queries such as ``CEO'' or ``scientist'').

With \textit{Journalist}, these interventions are likely to be implemented using semi-automated approaches, such as adding targeted re-ranking modules. For example, initial ranked results could be updated based on notions of \emph{fairness-aware ranking} \citep{ekstrand2022fairness,10.1145/3533379}; such modules are typically implemented using machine learning, and often trained using supervised methods (that is, using a set of human-generated labels from which the module learns to generalize). While these techniques are therefore initially ``tuned'' based on human input across example cases, they are then deployed in practice in an automated fashion without further intervention.

Importantly, a special case of ``censorship'' is carried out proactively by SEs: spam filtering. Web spam is a document that has no value for the user, but is intended to trick a SE to rank it highly (with e.g., the ultimate purpose of misleading a user into clicking on it and being led to undesirable content). SEs explicitly aim to filter out and remove spam from possible search results, based on the notion that such documents only have negative effects (i.e., making it harder for users to find information, while also consuming additional resources from search providers) and having no positive effects (i.e., they are by definition of no value to the user and simply designed to mislead).\footnote{Google provides extensive advice on what is considered to be spam, including the use of hidden text and links, keyword stuffing, and the buying and selling of links for ranking purposes (\url{https://developers.google.com/search/docs/essentials/spam-policies}, accessed 15 January 2024).}
Techniques for automated spam detection are typically based on statistical and machine learning approaches \citep{10.1145/1017074.1017077}. As a process that is automated at deployment time, spam filtering conceptually aligns with \textit{Journalist} (and perhaps \textit{Librarian}).

\textit{Model 4 (Teacher)} allows for even greater intervention in the best interest of the user and/or other stakeholders \citep{milano_recommender_2020}, including the possibility of explicitly prioritizing particular pieces of information, or, equally importantly, directly removing others considered to be harmful, trivial, or inappropriate. Again, practical SEs can already align with this model, and in practice may do so for a small number of queries, where direct human intervention can ``hardwire'' particular aspects of a search result page. A simple example could be an editor deciding that in response to the query ``covid cases today'', the top-ranked document must always be the advice page of the local country's health authority, no matter what ranking position other SE modules might have determined for this document.\footnote{Such interventions may also be carried out to comply with legal requirements, e.g., removal of illegal online material.}

Highly pertinent to \textit{Teacher} are recent developments in spoken conversational search, where queries are submitted using a voice channel and results returned as audio \citep{trippas2020model}. This has substantial implications for both the way in which queries are formed (e.g., it is much more common for users to formulate their queries using natural language or close approximations thereof, as opposed to simply listing a series of keywords) and for how results are presented (audio channel output will typically make it difficult to scan or browse items, and instead promotes linear consumption). Moreover, conversational search is an area of active development, where the search experience is moving from a query-response paradigm to one where an ongoing conversation is possible to facilitate more interactive and natural retrieval of information \citep{zamani2023conversational}. 

The very recent popularity of LLMs such as OpenAI's ChatGPT\footnote{\url{https://chat.openai.com}, accessed 15 January 2024.} have catapulted conversational information interaction into the mainstream, and LLM-based approaches have already been deployed into Microsoft's Bing \citep{mehdi2023reinventing} and Alphabet's Google \citep{reid2023generative} SEs.
The current behavior of some such systems is arguably aligned with the \textit{Teacher} model. For example, they provide direct answers through a conversational mechanism, and proactively select -- and ignore -- substantial subsets of available information from which an answer is generated. Moreover, in doing so they may seek to provide the user with expert information while steering away from trivial and morally problematic content. Of course, it is crucial to note that LLMs are not always successful in this \textit{Teacher} role. For instance, their training data and mode of operation (e.g. predicting most likely next words) can mean that they can hallucinate and produce poor information and ethically biased outputs. It would be ironic if such versions of \textit{Teacher} turned out to be a dangerous ``dilettante'' \cite{metzler2021rethinking}— less edifying and more unhelpful than existing SEs \cite{shah2022situating}. Even so, LLMs are improving to some extent in various dimensions, and the research area of generative information access (genIA) -- also referred as generative Information Retrieval (genIR) -- is advancing rapidly~\cite{white2025generative}.  We explore the impact of LLMs on the ethical considerations for SEs more fully in Section \ref{sec-LLMs}.

\section{COVID-19 Pandemic Case Study}
\label{sec:covid}

To illustrate how our role-based approach of four models could be practically applied, we discuss which models might have been justified during the pandemic. The way SEs responded to COVID-19 (mis)information was controversial \citep{ghezzi2020online,mehta_online_2021}. Interestingly, some studies showed the amount of vaccine misinformation from some SEs (e.g., Yahoo, Bing, Duckduckgo, Swisscows, Mojeek) significantly exceeded others (e.g., Google.com) \citep{ghezzi2020online}. In our case study of how the four models might apply to the pandemic, we do not definitively stake a position, but rather use the case study to further flesh out the models and illustrate how they could be employed.
As the physician-patient models do in medicine, our ethical models aid in reasoning about what a good SE might be for users and society. Again, there may be no single right model, and what is good might differ with the situation. Furthermore, precisely how the four models are \textit{used} will depend on what ethical or other theories, frameworks, reasons, etc. are invoked. For example, our discussion gives weight to personal autonomy \citep{beauchamp2001principles, young2017personal}and social liberty \citep{kymlicka2002contemporary, mill2023liberty}. Autonomy might be considered vital in liberal countries for various reasons, including a Kantian-like respect for moral agents who can set and follow their own ends \citep{kant2002groundwork}, and the contention that individuals tend to know better than the state and others, including corporations, what is good for them \citep{mill2023liberty}. Our case study also invokes an ethical duty to protect the community from significant harm \citep{brink_mills_2022}\footnote{Other values and arguments could be employed in this or in different cases -- our case study is simply an illustration of how the ethical models might be used, although it does contain arguments and reasons that might be extended to and deployed in other circumstances.} and illustrates how  "moral or political norms apply to search engines" \cite{munton2022answering}.

COVID-19 illustrated how vital SEs have become. People relied heavily on SEs to obtain crucial health information about vaccinations, lockdowns, masks, etc. \citep{mehta_online_2021}. A SE guided by the mere popularity of certain sites or the quantity of item views may propagate misinformation on a large scale. Popularity can but does not necessarily proxy for truth; indeed, political polarisation during COVID-19 allowed quack and conspiracy theories to flourish.

\myparagraph{Customer Servant and Librarian: User Autonomy}
To begin, it is surely perverse to say that SEs should ignore key aspects of the search query. A basic function of SEs is to provide information that people ask for, in a timely manner. So, if a queries COVID-19 vaccinations, the SE should return information about that and not unrelated subject matter. Given the internet is at least \textit{somewhat} like a public library, SEs should allow us to use that online resource efficiently and effectively.

However, it does not follow that \textit{Customer Servant} best respects autonomy, since the user may not clearly or explicitly state what they seek. Users may lack knowledge to ask the right questions or select appropriate keywords. Users also produce many different ``query variants'' that represent the same underlying information need; these different queries may produce search results that differ starkly in terms of effectiveness \citep{moffat_incorporating_2017}. For example, both ``closest vaccine centre'' and ``open covid clinic'' may be reasonable query attempts for an underlying information need to discover the location of the closest open COVID-19 vaccination clinic, but may produce substantially different result rankings on \textit{Customer Servant}.

\textit{Librarian}, which may use user click logs, personal interaction histories, resource popularity, etc., to infer intention, thus seems preferable to \textit{Customer Servant} here (and indeed in many other circumstances). It may be countered that inferring user intentions can be inaccurate (and paternalistic), and risks introducing biases from the system \citep{goldman2008searcha}. Still, if making inferences is generally more accurate, then respect for autonomy might favor \textit{Librarian}, all things being equal. Such respect during the pandemic was not only important for its own sake: many people, of course, wanted credible scientific information to protect themselves and their loved ones from serious harm.

\myparagraphdot{Journalist and Teacher: Protection or Oppression?}
\textit{Journalist} and \textit{Teacher} justify interfering with autonomy on the basis that certain information is in some sense harmful. \textit{Journalist} generally limits interference to changing result rankings. \textit{Teacher} goes further and may deny the user information considered harmful to them or society. Sometimes the impact on users may be similar, since most users only look at the top few answers \citep{spink_searching_2001}, but some differences may be salient. Imagine the query, ‘COVID-19 vaccination alters DNA in humans’. \textit{Teacher} might fill the first few pages not only with accurate and expert information about COVID-19 vaccinations, but also with sites that (say) explain the origin of conspiracy theories and vaccination panic, provide scholarly analyses of cognitive biases and rational thinking, etc. Of course, in the midst of the pandemic, accurate and comprehensive information about the virus (etc.) was frequently unavailable, and often reliable knowledge only emerged later, sometimes correcting earlier claims. Nonetheless, \textit{Teacher} (as an ideal type) could be expected to prioritize the best and most up-to-date information at the time, however incomplete or ultimately flawed it was later found to be.

Yet going beyond a user’s intentions to encourage deep understanding and foster intellectual or moral virtue will seem to many in liberal democracies too paternalistic. A supporter of \textit{Teacher} might object that it does not \textit{coerce} the user to consult that material. \textit{Teacher} might also provide some less edifying information, albeit well down the search rankings, and it only guides or 'nudges' \citep{thaler2021nudge} the user toward the most edifying links. 

But such nudging and re-directing still raises concerns about interference with liberty and autonomy \citep{schmidt_ethics_2020}. In addition to the fact that most users focus only on the top search results, many users may effectively treat SEs as possessing some testimonial authority \citep{narayanan2022google}. Down-ranking information may impede the user getting what they are seeking and thus may disrespect their autonomy. Moreover, \textit{Teacher} may filter out some information altogether, making it all but inaccessible.

Another possible problem with \textit{Teacher} is this. To acquire rounded and substantial knowledge of a subject, a student generally needs proper motivation. But arguably many or most internet searches are not motivated by a desire for deeper learning \citep{roseUnderstandingUserGoals2004}. For example, many who searched for COVID-19 information may have cared little for the precise pathophysiology, the history of virology, the in-depth political circumstances surrounding the outbreak, etc. They surely wanted to know many facts about the pandemic in considerable detail, but they were not generally seeking to acquire anything like expertise.

Furthermore, a deeper understanding of the pandemic was surely not required for protecting users or their loved ones. In some circumstances, it may be considered a justified act of paternalism to override autonomy to prevent great and irreversible harm from befalling a person. Such exceptions to respecting autonomy are debated in, for example, medical ethics \citep{beauchamp2001principles}. But \textit{Teacher} seems a disproportionate response for preventing such harms.

Could \textit{Journalist} be justified in (e.g.) a major health crisis despite a presumption in favor of user autonomy? \textit{Journalist} tries in part to give users the information they are seeking, but unlike \textit{Teacher} does not aim to promote deeper, more rounded knowledge or intellectual and moral virtue. Unlike \textit{Librarian}, \textit{Journalist} aims to educate by including some accurate information higher in the rankings, even when that entails demoting (though not to the same extent as \textit{Teacher}) some less accurate information the user may be looking for.

Consequently, some will argue that \textit{Journalist} wrongly interferes with autonomy and freedom. Some critics will call this objectionably paternalistic, even if it helps to prevent serious harm (e.g., life-threatening viral infection) to the user with relatively minimal intrusion upon their freedom. That may of course be debated. But for many, the goal of protecting society is a more compelling justification.

The vital role of decent journalism was highlighted during the pandemic when dangerous misinformation was amplified by technologies like social media algorithms \citep{saling2021no,spina2023human}. Since COVID-19 was pathogenic and contagious, even relatively few misinformed people could spread it with devastating results. And because SEs substantially affect the information and news we see and focus on, it seems reasonable to expect them to protect us from extremely harmful misinformation, even if they should concurrently restrict autonomy as little as possible. 
This position may recall John Stuart Mill’s harm principle \citep{mill2023liberty}.
On this much-discussed (though contested) principle \citep{holtug_harm_2002}, a person’s basic liberties may only be restricted to prevent substantial harm to others, not to further the good of that person. \citeauthor{mill2023liberty} envisages each person as free to pursue their own courses of action however bad or perverse they appear to others. \textit{Journalist} could conceivably respect this principle by interfering with liberty/autonomy only to protect others from significant harm. 

That said, the harm principle arguably opposes significant liberty reduction when the putative harm is relatively small and confined \citep{brink_mills_2022}. Some balance must be struck between impeded freedom and community protection. The right balance will often be vigorously contested, though disputants will at least share a moral understanding. Should some agreement be reached (e.g., via a fair procedure), the practical implications for a SE are various. These could range from ``brute force'' changes (e.g., automating the system so that any query that includes the term ``covid'' will always return the local country's health authority as the top ranked item), to more nuanced techniques that aim to provide further understanding about search intents (see Section~\ref{sec-se}). Additionally, the choice of technical solution would need to be informed by the risk appetite for false positives (e.g., still including some documents that convey misinformation towards the top of the rankings) versus false negatives (e.g., pushing some documents down in the rankings that do not in fact contain misinformation) -- another decision that may lead to conflicting views. The trade-offs described above are particularly more challenging for \emph{head queries} -- i.e., most popular queries issued to a SE -- where a SERP could be manually designed to include information obtained from knowledge graphs or specific databases. This was the case for the SERPs returned by commercial SEs for the query ``covid-19'' during the pandemic, which would include charts with real-time information about COVID-19 cases per region or country.

\myparagraph{Journalist and Teacher: Further Considerations}
An objection to \textit{Journalist} involves questioning the assumption that a person with (say) a certain search history and other characteristics available to the SE might believe and act on the misinformation. For example, some people may peruse fake health news and conspiracy theories out of curiosity, or even to guard against or debunk them. Here, the objection would be that liberty or autonomy are potentially hampered despite little risk of social harm. 
However, it is possible that while \textit{Journalist} will make some such mistakes, it will mostly be accurate about user intention regarding (say) COVID-19 misinformation. Yet a champion of non-interference may ask for empirical proof that SEs do cause the alleged social harms. But such effects will be complex and difficult to measure accurately (particularly as automation increasingly involves machine learning, where systems are trained on an initial data set and then set loose to generalize for new situations). Therefore, requiring ``proof'' may be asking too much when societies are plausibly at significant risk from online misinformation.

Another objection is that adopting \textit{Journalist} during global crises based on the harm principle creates a slippery slope. It is not only misinformation related to once-in-a-century pandemics that can cause social harm, but many other types of misinformation as well. This includes misinformation about history and politics that feeds distrust and intolerance. Some will embrace the apparent slippery slope, arguing that accurate information with some balance across a diversity of subject matter is essential to healthy democracy \citep{helberger_exposure_2018}. Indeed, such information is essential for the exercise of citizens’ political freedoms and autonomy—hence \textit{Journalist} should have widespread application. Others will counter that \textit{Librarian} suffices for protecting society from significant harms, such as damage to democracy from low-quality information. Citizens, they may argue, generally prefer accurate news, and democracy is anyway sturdy enough to tolerate any online ``filter bubbles'' \citep{michiels_what_2022,bruns2019are} and anti-democratic sentiment that algorithms may encourage.

However, threats to democracy are not the only reason for preferring \textit{Journalist}, or indeed \textit{Teacher}. Like other algorithms, SEs can produce biased results that are deeply prejudiced or socially damaging \citep{kulshresthaSearchBiasQuantification2019}. For example, a SE might present certain races, religious groups, genders, classes, or sexualities in offensive, hateful ways. Some may worry that recommendation algorithms will succor violent radicalization and extremism \citep{alfano_technological_2018}. Furthermore, algorithms might also seriously threaten the non-human world, such as by representing nonhuman animals as worthy of cruel treatment \citep{coghlan_harm_2023,hagendorff_speciesist_2023} or undermining social consensus about the urgent need to address environmental crises like climate change \citep{haider_google_2023}. 

In terms of our pandemic case study, it was apparent that skewed information and misinformation could bring additional harm and stigma to minority groups (e.g. poorer and marginalized people who sometimes suffered the brunt of the disease) \citep{jaiswal2020disinformation} and cause unnecessary harm to nonhuman animals who were assumed to be a zoonotic risk \citep{sellars2021one}. Because skewed, false, and hateful information may be very harmful, and may moreover sometimes be unjust in its own right, some will prefer elements of \textit{Journalist} or \textit{Teacher} to pure \textit{Librarian}. 

We might also note that up-ranking some accurate information at the expense of some socially harmful information could be done in varying degrees. While an SE might push all socially harmful information far down the rankings, it might alternatively leave some such information higher up while still downgrading some other amount, so as to strike a defensible balance between interfering with individual choice and protecting society. 

From an IR perspective, such an approach would require the automated identification of those documents that are ``socially harmful'' (since it is clearly not possible for human assessors to rate every document on the internet). Machine learning approaches have been successfully deployed for problems such as hate speech detection and sentiment analysis \citep{wankhadeSurveySentimentAnalysis2022}, and could feasibly be developed for this scenario. However, machine learning approaches raise their own challenges, including requiring carefully curated training examples (with the inherent complications of how one would define a ``social harmfulness'' scale on which documents would be rated, as well as who would ultimately decide what is or is not harmful), and the limited precision of effectiveness measurements when such a system is applied to previously unseen instances. Deploying any such approach for the re-ranking of search results will therefore also require the previously explained considerations regarding tradeoffs between false positives (where socially undesirable items are still incorrectly included in prominent positions in rankings) and false negatives (where items that are not in fact socially undesirable are incorrectly demoted). 

The COVID-19 case study shows how SEs play a pivotal role in situations where the spread of misinformation may cause serious community harm. We suggested how our four ethical models provide structure and clarity to certain trade-offs, including individual autonomy and community safety, and how these considerations relate to IR techniques. 

\section{Discussion}
\label{sec:discussion}

\subsection{Accountability of Private Platforms}

The four ethical SE models illustrate the varying levels of intervention that a SE can have on the information that is given to the user. The greater the degree of intervention—moving from \textit{Customer Servant} to \textit{Librarian} to \textit{Journalist} to \textit{Teacher}—the greater the ability of the SE to exert influence over the opinions and behavior of users. Thus, SE providers possess immense power to influence public opinion and societal behavior. 
This raises a crucial question about the accountability of SE providers to use their power in a responsible and ethical manner \citep{crawford_shortness_2005}. The most widely used SEs are owned and operated by private companies. They are subject to laws as any corporation is, but as private companies their primary motivation is to generate profits. Given the significant potential for SEs to influence the public, there is a risk that leaving this power in the hands of private companies motivated by profit could have undesirable consequences for society.
Recognizing the power of SEs to influence society, governments have responded by enacting laws that seek to regulate large SEs. For example, the European Union's Digital Markets Act \citep{eu2022dma} and Digital Services Act \citep{eu2022dsa} impose certain requirements on large SEs deemed ``gatekeepers''. These legal requirements come into effect in 2024, yet it is currently unclear how effective they will be in achieving their objectives. Governments are also responding by developing their own SEs, a key example being the European Open Web Index (OWI) \citep{granitzer2023impact} project.\footnote{\url{https://openwebsearch.eu/}, accessed 15 January 2024.} 

A positive factor about a State-owned and operated SE is that governments have obligations to act in the best interests of the community, and should not be motivated by profit. However, while democracy contains mechanisms to mitigate abuse of power, there is a risk that authorities may still seek to improperly influence the public through SEs. Government intervention could thus be politically motivated rather than genuinely in the community’s best interests. Hence, it is not clear that government-owned SEs are the best solution to the problem of accountability for SEs. 

Whether SEs are controlled by private companies or government authorities, the risk remains that any intervention can be used to undermine the autonomy of individuals, in ways that could be detrimental to individuals and society at large. But there is also a risk that insufficient intervention from SEs might do harm too. The four ethical models that we propose in this paper provide a role-based approach to discuss the challenges that SEs create for recognizing and balancing the autonomy of users and individual or wider harms. Further research is needed in the interdisciplinary areas of IR, ethics, and law to clarify the conditions under which it is ethical and legitimate for private companies or government authorities to deploy various SE models.

\subsection{The Prospect of Large Language Models for Search}
\label{sec-LLMs}

Finally, we consider the possible impact that LLMs will have in the near future on SEs. 
Practical LLMs such as OpenAI's ChatGPT, launched in late 2022, have popularized interactive information access, allowing users to engage in ``conversations'' by submitting natural language prompts to a system, and receiving natural language-like answers in response. LLMs are an example of \emph{generative} artificial intelligence (AI): they create new information, based on huge amounts of data on which they have previously been trained (e.g., including web pages, newspaper articles, encyclopedias, social media posts, and so on). These systems are remarkable in their ability to provide highly tailored responses for different user prompts, using their advanced capabilities for sentence completion \citep{bubeck2023sparks}. The technology has already been deployed in commercial SEs such as Microsoft's Bing Copilot
and Google's Generative Search Experience.
However, the use of LLMs brings several fundamental changes to information retrieval on the web \citep{shah2024envisioning,trippas2025adapting}. 

First, in line with their ``conversational'' interaction paradigm, these systems tend to produce a single answer at a time, consisting of a few sentences or paragraphs, rather than a ranked list of documents ordered by their expected utility for the user; answers are created by including -- and leaving out -- candidate information. 

Second, the answers are generated in response to a \emph{prompt}, based on what the model has previously been trained on, rather than directly extracting information from existing sources. LLMs do not explicitly model information (i.e., there is no underlying knowledge base), but rather are highly proficient in predicting ``next words'' based on millions of examples. This means that while generated responses are typically extremely well-formed natural language, the responses can be shallow or even factually incorrect (referred to as ``hallucinations''). Problematically, users of LLMs may not always be able to discern the value and veracity of the presented information \cite{Hicks2024}. 
Recent work by \citet{lajewska2024can} shows that factual correctness and query answerability are not easily identified by users in conversational information-seeking, not even when explainable mechanisms are incorporated~\cite{lajewska224explainability}. There is thus a concern about trust and trustworthiness for LLMs that does not apply to, say, competent human teachers.
 At the same time, improvements in LLMs may mean that they can be trusted to some extent, even when they are not infallible. Furthermore, other features might be implemented, such as designing LLM platforms that also give checkable links to reliable sources. However, LLM-based SEs will continue to be prone to errors -- as automatic information access systems -- and will perhaps never be as reliable as excellent human teachers.

Third, since the models are trained on vast quantities of data that may include toxic content, and are by nature not deterministic, they can provide problematic answers that range from being biased to being directly harmful or illegal. Therefore, these models are additionally trained with \emph{guardrails}, seeking to align the content that they generate with human values \citep{badyalIntentionalBiasesLLM2023}. A LLM that is presented with a prompt that would cause it to generate an answer that is at odds with its guardrails will typically result in a default response, informing the user that this topic is not appropriate. This raises concerns since these guardrails are not foolproof; indeed, a popular activity is the attempted ``jailbreaking'' of LLMs, aiming to trick the models into providing responses that they have been trained to avoid \citep{chaoJailbreakingBlackBox2023}. Guardrails are also potentially problematic even then they work as intended, as this is a direct form of intervention (and often censorship), aiming to restrict the information that can be provided to a user. Therefore, decisions about guardrails -- including what 
system behaviors are deemed ``inappropriate'' -- are key concerns. 

As a result of these considerations, we postulate that LLMs will substantially increase the level of intervention of SEs over users, and lead towards something like \textit{Teacher} becoming the default ethical model for LLM-based conversational SEs. As \citeauthor{shah2022situating} point out, LLM-based conversational SEs ``pose [an] even greater threat to transparency, provenance, and user interactions in a search system'' \citep{shah2022situating}. 

In contrast, SEs have previously implemented something resembling the \textit{Librarian} model for the majority of queries, where the SE has a low level of intervention. 
The move towards \emph{Teacher} represents a pronounced increase in the level of intervention that the SE exercises over access to information. The ability of SEs to influence public opinion and behavior, whether positive or negative, will likely be magnified by the increased level of intervention as we move from earlier \textit{Librarian} to future \textit{Teacher} models. 

It is possible that \textit{Teacher} could assist users in beneficial ways, including by developing their knowledge, insight, and even intellectual and ethical virtue. Further, LLM-based SEs must be very careful to avoid harmful biased outputs \citep{rozado_political_2023,ray_chatgpt_2023}. Hence, some substantial degree of influence over problematic search results may be partly and rightly unavoidable. But it is also possible that a\textit{ Teacher} SE designed to shape user knowledge and behavior in irresponsible ways could harm individuals, society as a whole, or the wider human and nonhuman world. Thus, apart from the issue of exercising some paternalism over information seekers, a further issue is that \textit{Teacher} might just be, for various reasons, bad at its job—just as some human teachers are bad at their jobs (though many are very good). Questions about the accountability of SEs, whether by private companies or government authorities, have ever greater urgency as LLMs shift SEs towards \textit{Teacher}-like models.

\subsection{Positionality and Ethical Considerations}

The authors span a range of disciplines, including Computer Science (with a focus on data science and information retrieval), Law, Moral Philosophy, and Digital Ethics. However, it is important to note the absence of expertise in other domains such as media studies and psychology. 
We acknowledge that our viewpoints come from a Western, liberal, democratic societal and cultural contexts, and that other perspectives (e.g., Indigenous perspectives \citep{yunkaporta2023right,lewis2020indigenous,janke2021true}) are not fully captured in our discussions.

The paper aims to bridge the gap between technology and ethics within the context of SEs. While there are opinions in the paper that reflect authors' viewpoint -- which may differ from those of the reader -- we hope the paper can facilitate further discussions across disciplines, contributing to a more comprehensive understanding around the ethical implications, regulation, and accountability associated with SEs.

\section{Conclusions and Further Work}
\label{sec:conclusions}

This paper examined what a ``good'' search engine might look like in an ethical sense involving users and other stakeholders, including society as a whole. We contributed an interdisciplinary analysis from the fields of ethics and information retrieval which offered a novel ethical role-based approach involving four search engine models: \textit{Customer Servant}, \textit{Librarian}, \textit{Journalist}, and \textit{Teacher}. The normative models are intended to clarify and sharpen understanding of how SEs have and might behave, just as the normative models of doctors did in medicine and medical ethics some decades ago. Each SE model has possible pros and cons, and there is not necessarily a single ``good'' search engine for all contexts. Our approach provides a foundation for further cross-disciplinary discussion of how ethical considerations can be translated and implemented using concepts in information retrieval. 

Via a COVID-19 case study, we showed how our four ethical models could be applied in a global health crisis and, to some extent, in other contexts. This exploratory exercise, which involved considerations of autonomy, liberty, harm, and social protection, illustrated how interested parties and stakeholders might critically discuss what type of search engine is best or justified in given circumstances. Further interdisciplinary research involving (say) law, IR, sociology, political science, etc. could develop our proposed approach and apply it in other contexts -- including business operations, verticals, and common business contexts. Finally, we highlighted the increasing urgency of such discussions at a time where new large language models (LLMs) may amplify the level of control that SEs have over online information and what information people see and do not see. 

In line with the emerging body of work that aims to characterize the societal implications of LLM-based conversational SEs~\citep{sakai2023few,shah2024envisioning,shah2023report,white2024cacm}, we believe the ethical models included in our proposed approach may contribute to a better understanding of the impact of SE technology in society.

\section*{Conflict of Interests}
 On behalf of all authors, the corresponding author states that there is no conflict of interest.

\section*{Disclosure and Acknowledgments}

Simon Coghlan (Senior Research Fellow) and Hui Xian Chia (PhD Student) are members of the Centre for AI and Digital Ethics (CAIDE). 
Falk Scholer and Damiano Spina are Associate Investigators of the ARC Centre of Excellence for Automated Decision-Making and Society (ADM+S).

The authors declare that did not make use of AI-assisted technologies. 

The authors would like to acknowledge Country. This research has been carried out on the unceded lands of the Woi Wurrung and Boon Wurrung language groups of the eastern Kulin nation. We pay our respects to their Ancestors and Elders, past, present, and emerging. We respectfully acknowledge their connection to land, waters, and sky.

This work was supported in part by the Australian Research Council (DP190101113, DE200100064) and the ARC Centre of Excellence for Automated Decision-Making and Society (CE200100005).

\section*{CRediT Authorship Contribution Statement}
\insertcreditsstatement

\bibliographystyle{cas-model2-names}

\bibliography{paper}

\end{document}